\documentclass[10pt,showpacs,preprintnumbers,amsmath,amssymb,amsthm]{revtex4}
\usepackage{units,hyperref,fancyhdr}
\usepackage{bbold,mathpazo}
\usepackage{graphicx}
\usepackage{dcolumn}
\usepackage{bm,bbm,mathtools,esvect}
\usepackage{slashed}
\begin{document}

\title{Quaternionic  Quantum  Particles: New Solutions}
\vspace{2cm}
\author{ SERGIO GIARDINO}
\email{giardino.sergio@unifesp.br}
\affiliation{\vspace{3mm} Institute of Science and Technology, Federal University of S\~ao Paulo\\
Avenida Cesare G. M. Lattes 1201, 12247-014 S\~ao Jos\'e dos Campos, SP, Brazil}

\begin{abstract}
\noindent If $\Psi$ is a quaternionic wave function, then $i\Psi\neq \Psi i$. Thus, there are two versions of
the quaternionic Schr\"odinger equation (QSE). In this article, we present the second possibility for solving the QSE, following on from
a previous article. After developing the general methodology, we present the
quaternionic free particle solution and the scattering of the quaternionic particle through a scalar barrier.
\end{abstract}

\maketitle
\hrule
{\parskip - 0.3mm \footnotesize{\tableofcontents}}
\vspace{1cm}
\hrule

\section{INTRODUCTION\label{I}}%
Quaternions \cite{Rocha:2013qtt}, represented by $\mathbb{H}$, are hyper-complex numbers with three anti-commuting complex units. In general, if 
$q\in\mathbb{H}$, then
\begin{equation}\label{I0}
 q=x_0+x_1 i+x_2 j+ x_3 k,
\end{equation}
where $x_0,\,x_1,\,x_2$ and $ x_3$ are real and the complex units $i,\;j$ and $k$ satisfy
\begin{equation}\label{I1}
ij=-ji=k\qquad \mbox{and}\qquad ijk=-1.
\end{equation}
Using a notation called symplectic, we express (\ref{I0}) as
\begin{equation}\label{I2}
 q=z +\zeta j\qquad\mbox{with}\qquad z,\,\zeta\in\mathbb{C}.
\end{equation}
Quantum mechanics is a physical theory based on complex numbers. A generalization that replaces complexes with quaternions
has been attempted, and Stephen Adler's textbook \cite{Adler:1995qqm} surveys this endeavor. In this formulation, quaternionic quantum
mechanics (QQM) deploys anti-hermitian operators (AHO), instead of the usual hermitian operators of complex quantum mechanics (CQM). 
An anti-hermitian operator $\mathcal{A}$ satisfies $\mathcal{A}^\dagger=-\mathcal{A}$, where $\mathcal{A}^\dagger$ is the adjoint operator of 
$\mathcal{A}$. QQM has been formulated in terms of AHO, among other reasons, to preserve the conservation of the probability current. 
In spite of this advantage, the adoption of AHO has several drawbacks. The most visible is the breakdown of Ehrenfest's theorem
\cite{Adler:1985uh,Adler:1988fs,Adler:1995qqm}. The lack of simple solutions in anti-hermitian QQM is another drawback.
Of course, there are anti-hermitian quaternionic solutions, like
\cite{Davies:1989zza,Davies:1992oqq,Ducati:2001qo,Nishi:2002qd,DeLeo:2005bs,Madureira:2006qps,Ducati:2007wp,Davies:1990pm,DeLeo:2013xfa,DeLeo:2015hza,Giardino:2015iia,Sobhani:2016qdp,Procopio:2016qqq},
but they are difficult to grasp and do not have the simplicity that allows them to be compared to either CQM  or classical solutions. 
Experimental tests have been performed as well \cite{Walter:2017prw,Brumby:1996xf}, but no quaternionic effect has been observed at the present time

More recently, novel alternatives have been attemptei to build a consistent QQM. One may replace usual QQM wave functions with regular 
quaternionic
functions \cite{Sabadini:2017wha}, and maintain anti-hermitian formalism. A more radical approach emerged after the discovery of quaternionic 
solutions obtained throughout non-anti-hermitian (NAH) Hamiltonians in the study of the quaternionic Aharonov-Bohm (AB) effect \cite{Giardino:2016xap}.
This discovery has enabled a formal expression of an NAH-QQM, where the probability current and the expectation value are redefined \cite{Giardino:2017nah}.
Following these results, a solution for the QSE was developed, and the first quaternionic particle solution
was obtained \cite{Giardino:2017qqp}. 

Because of the non commutativity between quaternionic functions and the complex unit $i$, there are two 
possibilities for the QSE, according to the position of $i$. The first possibility has already been considered in \cite{Giardino:2017qqp}, and
in this article we entertain the second possibility. 

This article is organized as follows: Section \ref{A} presents a general time-dependent solution for QSE, whereas the
time-independent QSE is solved in Section \ref{B}. In Section \ref{T}, the solution method is used in several simple situations and, in
Section \ref{F}, the quaternionic free particle is obtained. Section \ref{S} describes the scattering of the quaternionic free particle 
through a scalar step potential, while section \ref{C} rounds off the article with our conclusions and future perspectives.

\section{TIME-DEPENDENT EQUATION\label{A}}

The time-dependent quaternionic Schr\"odinger is 
\begin{equation}\label{A1}
i\, \hbar\,\frac{\partial\Psi}{\partial t}=\mathcal{H}\Psi,
\end{equation}
where $\Psi$ is a quaternionic wave function and $\mathcal{H}$ is a quaternionic Hamiltonian.
We comment on the position of the complex unit $i$ on the {\it left} side of the time-derivative of $\Psi$ in (\ref{A1}). We separate the time variable of the wave
function as
\begin{equation}\label{A2}
 \Psi(\bm x,\,t)=\Phi(\bm x)\Lambda(t),
\end{equation}
where $\Phi(\bm x)$ and $\Lambda(t)$ are both quaternionic. In a complex wave function, a complex exponential,
which is a unitary complex, carries the time dependence. By analogy, the time-dependent function $\Lambda$ of the quaternionic wave function
will be chosen to be a unitary quaternionic function. Accordingly, in symplectic notation (\ref{I2}), we get
\begin{equation}\label{A3}
 \Lambda=\cos\Xi\, e^{iX}+\sin\Xi\, e^{i \Upsilon}j,\qquad\mbox{so that}\qquad \Lambda\Lambda^*=1.
\end{equation}
$\Xi,\,X$ and $\Upsilon $ are time-dependent real functions and $\Lambda^*$ is the quaternionic conjugate of $\Lambda$.
In order to eliminate the dependence on time from the wave equation, we multiply the right hand side of (\ref{A1}) by $\Lambda^*$ and 
impose
\begin{equation}\label{A5}
 \dot\Lambda \Lambda^*=\frac{\kappa}{\hbar},
\end{equation}
where the dot denotes a time derivative and $\kappa\in\mathbb{H}$ is the separation constant
\begin{equation}
\kappa=\kappa_0+\kappa_1 j,\qquad\mbox{with}\qquad \kappa_0,\,\kappa_1\in\mathbb{C}.
\end{equation}
 Consequently, from (\ref{A5}) we get
\begin{equation}\label{A6}
i\Big( \dot X\cos^2\Xi+\dot \Upsilon\sin^2\Xi\Big)+\Big[\dot\Xi+i\sin\Xi\cos\Xi\big(\dot \Upsilon-\dot X\big)\Big]e^{i(X+Y)}j=\frac{\kappa}{\hbar}.
\end{equation}
We can obtain several solutions for (\ref{A6}). For $\,\dot\Xi=0,\,\dot X=\dot \Upsilon=-i\kappa_0$ and $\,\kappa_1=0$, we
obtain
\begin{equation}\label{A7}
 \Lambda=\exp\left[-\frac{i\mathcal{E}}{\hbar}t\right]\Lambda_0,\qquad\mbox{so that}\qquad\dot\Lambda\Lambda^*=-\frac{i\mathcal{E}}{\hbar},
\end{equation}
where $\mathcal{E}$ is the energy and $\Lambda_0$ is a unitary quaternionic constant that can multiply both sides of the
complex exponential indifferently. The time solution (\ref{A7}) is very similar to the complex case; conversely, we achieved
something more interesting by imposing $\,\dot\Xi=0$ and $ X+ \Upsilon=\tau_0$, where $\tau_0$ is a real constant. Thus, we get
\begin{equation}\label{A8}
\Lambda=\left\{\cos\Xi\exp\left[-\frac{i\mathcal{E}}{\hbar}t\right]+\sin\Xi\,\exp\left[i\left(\frac{\mathcal{E}}{\hbar}t+\tau_0\right)\right]j\right\}\Lambda_0,\qquad\mbox{so that}\qquad
\dot\Lambda\Lambda^*=\frac{i\mathcal{E}}{\hbar}\left(-\cos2\Xi+\sin2\Xi\,e^{i\tau_0}j\right).
\end{equation}
We remark that (\ref{A8}) may be written schematically as
\begin{equation}\label{A9}
 \Lambda=\Lambda_1\exp\left[-\frac{i\mathcal{E}}{\hbar}t\right]\Lambda_0
\end{equation}
where $\Lambda_1$ is an arbitrary constant quatenion. Solution (\ref{A8}) is absolutely new, it recovers (\ref{A7}) when $\sin\Xi=0$, but
the value of the constant quaternion contributes to the eigenvalue. This kind of influence of a constant over the eigenvalue of an eigenfucntion
is unknown in CQM. We expect that novel quantum solutions, unattainable through CQM may 
emerge. Nevertheless, there is another simple possibility for time-dependent solutions. If $\dot X=\dot \Upsilon=0$, we achieve
\begin{equation}\label{A10}
\Lambda=\left[\cos\left(\frac{\mathcal{E}}{\hbar}t\right)e^{-iX}+
\sin\left(\frac{\mathcal{E}}{\hbar}t\right)e^{i\left(X+\tau_0\right)}j\right]\Lambda_0\qquad\mbox{so that}\qquad
 \dot\Lambda\Lambda^*=\frac{\mathcal{E}}{\hbar}\,e^{i\tau_0}\,j.
\end{equation}
In the same fashion as (\ref{A8}), the solution given by (\ref{A9}) is absolutely new, it does not have a complex counterpart and
a complex limit is meaningless. If something physical may be described with it, it is probably unknown to CQM.

\section{GENERAL TIME-INDEPENDENT EQUATION\label{B}}

Using (\ref{A1}-\ref{A5}), the time-independent Schr\"odinger equation reads
\begin{equation}\label{B1}
 \mathcal{H}\Phi=i\,\Phi\,\kappa.
\end{equation}
In order to solve (\ref{B1}), we need several assumptions. First of all, we propose the spatial wave function
\begin{equation}\label{B2}
\Phi=\phi\,\lambda,
\end{equation}
where $\lambda$ is a time-independent quaternionic function given by
\begin{equation}\label{B4}
 \lambda=\rho K,\qquad\mbox{where}\qquad |\lambda|=\rho,\qquad K=\cos\Theta\, e^{i\Gamma}+\sin\Theta\, e^{i \Omega}j\,,
\qquad\mbox{and}\qquad KK^*=1;
\end{equation}
and $\phi$ is a time-independent complex solution of Schr\"odinger equation with energy $E$, so that 
\begin{equation}\label{B3}
 \mathcal{H}\phi=E\phi,
\end{equation} 
where the energy $E$ is real. Consequently, the Hamiltonian $\mathcal{H}$ is the hermitian operator
\begin{equation}\label{B5}
 \mathcal{H}=-\frac{\,\hbar^2}{2m}\nabla^2+V,
\end{equation}
where $V$ is a real scalar potential. More general hamiltonian operators, with complex potentials and complex energies, 
are interesting directions for research and they may be examined in a separate article. Using (\ref{B1}-\ref{B2}), (\ref{B4}), we get
\begin{equation}\label{B6}
\nabla^2\lambda+\frac{2}{\phi}\bm\nabla\phi\bm{\cdot\nabla}\lambda=\frac{2m}{\hbar^2}\big(E\lambda -i\lambda \kappa\big).
\end{equation}
We also adopt
\begin{equation}\label{B7}
 \bm\nabla K=\bm p\, e^{i\Gamma}+\bm q\, e^{i\Omega}j\qquad\mbox{and}\qquad \nabla^2 K=u\, e^{i\Gamma}+ v\, e^{i\Omega}j,
\end{equation}
where
\begin{align}
& \bm p=-\sin\Theta\,\bm\nabla\Theta+i\cos\Theta\,\bm\nabla\Gamma,\qquad
\bm q= \cos\Theta\,\bm\nabla\Theta+i\,\sin\Theta\,\bm\nabla\Omega,\nonumber\\
& u= -\cos\Theta\,\Big(\,\big|\bm\nabla\Gamma\big|^2+\big|\bm\nabla\Theta\big|^2\,\Big)-\sin\Theta\,\nabla^2\Theta
+i\Big(\cos\Theta\,\nabla^2\Gamma-2\sin\Theta\,\bm\nabla\Gamma\bm{\cdot\nabla}\Theta\Big)\label{B8}\\
 & v=-\sin\Theta\,\Big(\,\big|\bm\nabla\Omega\big|^2+\big|\bm\nabla\Theta\big|^2\,\Big)+\cos\Theta\,\nabla^2\Theta
+i\Big(\sin\Theta\,\nabla^2\Omega+2\cos\Theta\,\bm\nabla\Omega\bm{\cdot\nabla}\Theta\Big).\nonumber
\end{align}
The complex and quaternionic parts of (\ref{B6}) give
\begin{align}
& \label{B9}
\frac{1}{\rho}\left(\bm\nabla+\frac{2}{\phi}\bm\nabla\phi\right)\bm{\cdot\nabla}\rho+
\frac{2}{\rho\phi}\bm\nabla(\rho\phi)\bm\cdot\frac{\bm p}{\cos\Theta}+\frac{u}{\cos\Theta}=
\frac{2m}{\hbar^2}\left[E-i\kappa_0+i\kappa^*_1\tan\Theta \,e^{i(\Omega-\Gamma)}\right] \\
& \label{B10}
\frac{1}{\rho}\left(\bm\nabla+\frac{2}{\phi}\bm\nabla\phi\right)\bm{\cdot\nabla}\rho+
\frac{2}{\rho\phi}\bm\nabla(\rho\phi)\bm\cdot\frac{\bm q}{\sin\Theta}+\frac{v}{\sin\Theta}=
\frac{2m}{\hbar^2}\left[E-i\kappa_0^*-i\kappa_1\cot\Theta \,e^{i(\Gamma-\Omega)}\right].
\end{align}
Specific values chosen for $\kappa$ furnish the three time-dependent solutions (\ref{A7}-\ref{A9}). Four real equations
are obtained from the two complex equations (\ref{B9}-\ref{B10}). First of all, we note that
\begin{equation}\label{B11}
i\kappa_0\in\mathbb{R}\qquad\mbox{and}\qquad \kappa_1=|\kappa_1|e^{i\tau_0}.
\end{equation}
Thence, we define
\begin{equation}\label{B12}
 \frac{1}{\rho}\left(\bm\nabla+\frac{2}{\phi}\bm\nabla\phi\right)\bm{\cdot\nabla}\rho=\mathcal{Z}_0,\qquad
\frac{2}{\rho\phi}\bm\nabla(\rho\phi)\bm\cdot\frac{\bm p}{\cos\Theta}=\mathcal{Z}_1\qquad\mbox{and}\qquad
\frac{2}{\rho\phi}\bm\nabla(\rho\phi)\bm\cdot\frac{\bm q}{\sin\Theta}=\mathcal{Z}_2,
\end{equation}
where $\mathcal{Z}_0,\,\mathcal{Z}_1$ and $\mathcal{Z}_2$ are complex functions.
We finally separate (\ref{B9}-\ref{B10}) into real components, so that
\begin{align}
\label{B13}
& \Re(\mathcal{Z}_0+\mathcal{Z}_1)-|\bm\nabla\Gamma\big|^2-\big|\bm\nabla\Theta\big|^2-\tan\Theta\nabla^2\Theta
=\,\frac{2m}{\hbar^2}\Big(E-i\kappa_0+|\kappa_1|\tan\Theta\sin W\Big) \\
\label{B14}
& \Re(\mathcal{Z}_0+\mathcal{Z}_2)-|\bm\nabla\Omega\big|^2-\big|\bm\nabla\Theta\big|^2+
\cot\Theta\nabla^2\Theta=\,\frac{2m}{\hbar^2}\Big(E+i\kappa_0+|\kappa_1|\cot\Theta\sin W\Big) \\
\label{B15}
&\Im(\mathcal{Z}_0+\mathcal{Z}_1)+\Big(\bm\nabla-2\tan\Theta\bm\nabla\Theta\Big)\bm{\cdot\nabla}\Gamma=\,\frac{2m}{\hbar^2}|\kappa_1|\tan\Theta\cos W\\
\label{B16}
&\Im(\mathcal{Z}_0+\mathcal{Z}_2)+\Big(\bm\nabla+2\cot\Theta\bm\nabla\Theta\Big)\bm{\cdot\nabla}\Omega=-\frac{2m}{\hbar^2}|\kappa_1|\cot\Theta\cos W
\end{align}
where 
\begin{equation}  \label{B17}
 W=\Gamma-\Omega+\tau_0, 
\end{equation}
and $\Re(\mathcal{Z})$ and $\Im(\mathcal{Z})$ are the real and the imaginary components of a complex $\mathcal{Z}$, respectively. 
Equations (\ref{B13}-\ref{B16}) are the most general time-independent solutions obtained from the three time-dependent cases. 
In the following section, we examine several simple solutions, and leave more complicated cases for future research.

\section{WAVE FUNCTIONS\label{T}}

Assuming an arbitrary one-dimensional complex wave function $\phi$ and also $\bm\nabla\Theta=\bm 0$, we impose the  constraints 
\begin{equation}\label{T0}
\bm\nabla\phi\bm{\cdot\nabla}\rho=0,\qquad \bm\nabla\phi\bm{\cdot\nabla}\Gamma=0,\qquad \bm\nabla\phi\bm{\cdot\nabla}\Omega=0,\qquad
\bm\nabla\rho\bm{\cdot\nabla}\Gamma=0,\qquad\bm\nabla\rho\bm{\cdot\nabla}\Omega=0,
\end{equation}
remembering that $\bm\nabla\Gamma\bm{\cdot\nabla}\Omega\neq 0$. Thus, (\ref{B13}-\ref{B16}) turn into two equations
\begin{align}
\label{T1}
&\frac{1}{\rho}\nabla^2\rho-|\bm\nabla\Gamma|^2=\frac{2m}{\hbar^2}\Big(E-i\kappa_0+|\kappa_1|\tan\Theta\sin W\Big)\\
\label{T2}
&\frac{1}{\rho}\nabla^2\rho-|\bm\nabla\Omega|^2=\frac{2m}{\hbar^2}\Big(E+i\kappa_0+|\kappa_1|\cot\Theta\sin W\Big),\\
\label{T3}
&\nabla^2\Gamma=\,\frac{2m}{\hbar^2}|\kappa_1|\tan\Theta\cos W\\
\label{T4}
&\nabla^2\Omega=-\frac{2m}{\hbar^2}|\kappa_1|\cot\Theta\cos W
\end{align}
From (\ref{T0}), we suppose that there is no common variable between either $\rho$ and $\Gamma$ or between $\rho$ and $\Omega$;
therefore, $\nabla^2\rho/\rho$ has to be constant. After the action of the gradient operator over (\ref{T1}-\ref{T2}), 
we recover (\ref{T3}-\ref{T4}) with changed signs. This result imposes
\begin{equation}\label{T5}
 \nabla^2\Gamma=\nabla^2\Omega=0.
\end{equation}
Hence, we gain the constraint
\begin{equation}\label{T6}
 |\kappa_1|\cos W=0,
\end{equation}
and consequently two cases to consider.  Let us examine the first one.

\subsection{$|\kappa_1|=0$\label{k1}}

We suppose a three-dimensional space, and thus $\bm\nabla\Gamma$ and $\bm\nabla\Omega$ are necessarily collinear.
The other two directions of the space given by $\bm\nabla\phi$ and $\bm\nabla\rho$. From (\ref{T1}) and (\ref{T2}), we obtain
\begin{equation}\label{T7}
\frac{1}{\rho}\nabla^2\rho=\frac{2mE}{\hbar^2}+\frac{|\bm\nabla\Gamma|^2+|\bm\nabla\Omega|^2}{2},\qquad\mbox{and}\qquad 
\frac{2m\mathcal{E}}{\hbar^2}=\frac{|\bm\nabla\Gamma|^2-|\bm\nabla\Omega|^2}{2},
\end{equation}
where 
\begin{equation}\label{T8}
\mathcal{E}=i\kappa_0
\end{equation}
 is the quaternionic energy. Therefore, we reach the following solution
\begin{equation}\label{T9}
 \Gamma=\bm{\gamma\cdot x}+\Gamma^{(0)},\qquad \Omega=\bm{\omega\cdot x}+\Omega^{(0)},\qquad\mbox{and}\qquad 
\rho=A\, e^{\,\bm{\alpha\cdot x}}+B\,e^{-\bm{\alpha\cdot x}},
\end{equation}
where $\Gamma^{(0)},\,\Omega^{(0)},\,A$ and $B$ are real scalar constants, and $\bm\alpha$ is a constant real vector. There is no solution 
for the constant $\rho$, and we must have 
\begin{equation}\label{T10}
 |\bm\alpha|^2=\frac{2mE}{\hbar^2}+\frac{|\bm\gamma|^2+|\bm\omega|^2}{2}\qquad\mbox{and}\qquad|\bm\gamma|^2-|\bm\omega|^2\geq 0.
\end{equation}
The most general wave function is thus
\begin{align}\nonumber
& \Phi=
\phi(\bm x)\,\rho(\bm x)\Big[\left(\cos\Theta\, e^{\,i\Gamma}+\sin\Theta\, e^{\,i\Omega}j\right)C_1+
\left(\cos\Theta\,e^{\,i\Gamma}+\sin\Theta\,e^{-i\Omega}j\right)C_2+\\
&\label{T11}
+\left(\cos\Theta\, e^{-i\Gamma}+\sin\Theta\, e^{\,i\Omega}j\right)C_3+
\left(\cos\Theta\,e^{-i\Gamma}+\sin\Theta\,e^{-i\Omega}j\right)C_4
\Big],
\end{align}
where $C_1,\,C_2,\,C_3$ and $C_4$ are arbitrary complex constants. (\ref{B1}) is not satisfied for quaternionic integration constants. 
We can make $\phi$ constant, so that $E=0$, and thus obtain the simplest solution of the case, a truly quaternionic free particle. 
However, we stress that every one-dimensional complex wave function $\phi$ generates the same kind of quaternionic solution, where
the quaternionic solution may be understood as a geometric phase. A general study concerning quaternionic phases is an interesting
direction for research.

\subsection{$\cos W= 0$ and $\kappa_0\neq 0$}
The solutions of this case obey
\begin{equation}\label{T12}
W= \Gamma-\Omega+\tau_0=\left(n+\frac{1}{2}\right)\pi\qquad\mbox{so that}\qquad\bm\nabla\Gamma=\bm\nabla\Omega,
\end{equation}
with $n\in\mathbb{Z}$. Using (\ref{T1}-\ref{T2}), (\ref{T8}) and (\ref{T12}), we get
\begin{equation}\label{T13}
 \mathcal{E}=-|\kappa_1|\cot 2\Theta\sin W,
\end{equation}
so that 
\begin{equation}\label{T14}
\frac{1}{\rho}\nabla^2\rho=\frac{2m}{\hbar^2}\Big(E-\mathcal{E}\sec2\Theta\Big)+|\bm\nabla\Gamma|^2.
\end{equation}
Inasmuch as there is no defined sign on the right hand side of (\ref{T14}), two kinds of solutions for $\rho$ are admitted, 
either real exponentials or a linear combination of sines and cosines. Even a $\nabla^2\rho=0$ is admitted, and hence there are
more possibilities for $\rho$ in this situation than has been found in the previous $|\kappa_1|=0$ case.
The general solution is thus
\begin{equation}\label{T15}
\Phi=
\phi(\bm x)\,\rho(\bm x)\left( C_1\, e^{i\Gamma}+C_2\, e^{-i\Gamma}\right)\left(\cos\Theta-i\sin W \sin\Theta e^{i\tau_0}\,j\right)
\end{equation}
with $C_1$ and $C_2$ arbitrary complex constants and $\Gamma$ given by (\ref{T9}). The solution (\ref{T15}) comprises a complex
solution and a quaternionic unitary constant that multiplies its right-hand side. 
The quaternionic energy accomplished through (\ref{T14}) is
\begin{equation}\label{T16}
 \mathcal{E}=\cos2\Theta\left[E+\frac{\hbar^2}{2m}\left(|\bm\gamma|^2\pm|\bm\alpha|^2\right)\right],
\end{equation}
where the sign of $|\bm\alpha|^2$ is defined by $\rho$. It is totally unexpected that the
quaternionic constant influences the energy of the system, and there is no counterpart to this phenomenon in CQM. A
physical system described with this solution would be remarkable.
We notice that low quaternionic energies, where $\mathcal{E}<E$, are admitted, depending on $\rho$. 
This curious and new effect is probably due to the quaternionic character of the eigenvalue $\kappa$, which is connected to the energy through
(\ref{T13}). 

\subsection{$\cos W= 0$ and $\kappa_0 = 0$}

This solution is related to the time-dependent solution (\ref{A10}). The solutions of this case follow (\ref{T12}), but (\ref{T1}-\ref{T2})
additionally imposes
\begin{equation}\label{T17}
 \Theta=\left(\tilde n+\frac{1}{2}\right)\frac{\pi}{2},\qquad\mbox{where}\qquad \tilde n\in\mathbb{Z}.
\end{equation}
Consequently, the quaternionic wave function is obtained from (\ref{T15}), and the quaternionic energy is
\begin{equation}\label{T18}
\mathcal{E}=|\kappa_1|=-\frac{\cot\Theta}{\sin W} \left[E+\frac{\hbar^2}{2m}\left(|\bm\gamma|^2\pm|\bm\alpha|^2\right)\right].
\end{equation}

\subsection{The $\pmb\nabla\Gamma=\pmb\nabla\Omega=\pmb 0\;$ case\label{G}}

Assuming an arbitrary one-dimensional complex wave function $\phi$, we impose the  constraints 
\begin{equation}\label{G0}
\bm\nabla\phi\bm{\cdot\nabla}\rho=0,\qquad \bm\nabla\phi\bm{\cdot\nabla}\Theta=0,\qquad \bm\nabla\rho\bm{\cdot\nabla}\Theta=0.
\end{equation}
Thus, (\ref{B12}-\ref{B15}) turns into two equations
\begin{align}
\label{G1}
&\frac{1}{\rho}\nabla^2\rho-|\bm\nabla\Theta|^2=\frac{2m}{\hbar^2}\Big(E-i\kappa_0\cos2\Theta+|\kappa_1|\sin W \sin2\Theta\Big)\\
\label{G2}
&\nabla^2\Theta=\frac{2m}{\hbar^2}\Big(i\kappa_0\sin2\Theta+|\kappa_1|\sin W\cos2\Theta\Big), 
\end{align}
and the constraint (\ref{T6}) that has already been found for the $\bm\nabla\Theta=\bm 0$ cases. However, the orthogonality
of $\bm\nabla\rho$ and $\bm\nabla\Theta$ implies that $\nabla^2\rho/\rho$ is a constant. Applying the gradient operator over
(\ref{G1}), we recover (\ref{G2}) with a changed sign, and thus $\nabla^2\Theta=0$, which forces a constant
$\Theta$ because of (\ref{G2}). With exception for an exotic zero energy solution, where $\kappa=0$ and $\bm\nabla\Theta$ is constant. 
Therefore, we do not have a simple solution for non-constant $\Theta$. Maybe this kind of
solution exists when (\ref{G0}) includes a non-orthogonal $\bm\nabla\Theta$. The research of such solutions is potentially an
interesting subject for future work.

\section{THE FREE QUATERNIONIC PARTICLE\label{F}}

Let us take the complex free particle as a reference. Its expression is
\begin{equation}\label{F0}
\phi(\bm x)=A_1\,e^{\,i\bm{k\cdot x}}+A_2\,e^{-i\bm{k\cdot x}},\qquad\mbox{where}\qquad |\bm k|^2=\frac{2mE}{\hbar^2},
\end{equation}
and $A$ and $B$ are complex integration constants. In order to understand the physics of the previously calculated solutions, we will 
entertain the probability current defined in \cite{Giardino:2016xap,Giardino:2017nah}, namely 
\begin{equation}\label{F1}
\bm j=\frac{1}{2m}\left\{\Big(\hat{\bm p}\Psi\Big)\Psi^*+\left[\Big(\hat{\bm p}\Psi\Big)\Psi^*\right]^*\,\right\}\qquad\mbox{and}\qquad
\hat{\bm p}\Psi=-i\,\hbar\bm\nabla\Psi.
\end{equation}
Using the general quaternionic time independent wave function (\ref{B2}), we get
\begin{equation}\label{F2}
\bm j=\rho^2\left[\;\bm j_0+\frac{\hbar}{m}|\phi|^2\left(\cos^2\Theta\bm\nabla\Gamma+\sin^2\Theta\bm\nabla\Omega\right)\right] |C|^2
\end{equation}
where $C$ is an arbitrary quaternionic constant and $\bm j_0$ is the probability current due to the complex free particle (\ref{F0}). 
There is no probability flux along the $\bm \nabla\Theta$ and $\bm\nabla\rho$ directions. This fact does not mean that there is no motion 
along these directions. We remember the complex square well, where a particle oscillates along a confined region without generating a
non-zero probability flux. Thus, we interpret that the quaternionic particle freely propagates along directions 
$\bm\nabla\Gamma$ and $\bm\nabla\Omega$ only, while directions $\bm \nabla\Theta$ and $\bm\nabla\rho$ allow oscillatory motions only.
A simple quaternionic free particle obtained from (\ref{T11}) is 
\begin{equation}\label{F3}
\Phi=\rho(\bm x)\left(\cos\Theta\, e^{i\bm{\gamma\cdot x}}+\sin\Theta\, e^{i\bm{\omega\cdot x}}\,j\right)C\qquad\qquad\Rightarrow\qquad\qquad 
\bm j= \frac{\hbar}{m}\rho^2|C|^2\left(\cos^2\Theta\,\bm\gamma+\sin^2\Theta\,\bm\omega\right),
\end{equation}
where $C$ is a quaternionic constant and  $\bm\nabla\Theta=\bm 0$. From (\ref{T7}), we get
\begin{equation}\label{F4}
|\bm\alpha|^2=\frac{|\bm\gamma|^2+|\bm\omega|^2}{2},\qquad\qquad 
\frac{2m\mathcal{E}}{\hbar^2}=\frac{|\bm\gamma|^2-|\bm\omega|^2}{2}.
\end{equation}
An exotic zero-energy wave function is possible for
\begin{equation}\label{F5}
 |\bm\alpha|=|\bm\gamma|=|\bm\omega|.
\end{equation}
The situation is different from the complex case, where null energies imply null momenta. 
Another particularity is that we cannot recover a complex solution by simply imposing $\Theta=|\bm\omega|=|\bm\alpha|=0$. 
 We remember that $|\bm\alpha|=0$ is prohibited by (\ref{F4}), and that a simple quaternionic constant may change the energy
of the particle (\ref{T15}-\ref{T16}). Thus, QQM solutions cannot be understood as encompassing a complex
solution plus an independent pure quaternionic part. In fact, we have a different theory, which may recover CQM for several situations, 
but not for all cases.

\section{\label{S}THE STEP POTENTIAL}

Let us consider the scalar step potential
\begin{equation}\label{S0}
 V=\left\{
\begin{array}{ccc}
0 & \qquad\; \mbox{for}\qquad  x<0,&\;\;\;\;\;\mbox{region I} \\
V_0 & \qquad \mbox{for}\qquad  x\ge 0, &\qquad\mbox{region II},
\end{array}
\right.
\end{equation}
where $V_0$ concerns a real positive constant  and the potential $V$ divides the three-dimensional space into
two parts bordered by the $Oyz$ plane. We propose the wave function
\begin{align}\nonumber
\Phi_I&=\rho_k\left[\cos\Theta_k e^{i(\bm k+\bm\gamma_k^\perp)\bm{\cdot x}}+\sin\Theta_k e^{i(-\bm k+\bm\omega_k^\perp)\bm{\cdot x}}\,j\,\right]
+ R\,\rho_q\left[\cos\Theta_q e^{i(-\bm q+\bm\gamma_q^\perp)\bm{\cdot x}}+\sin\Theta_q e^{i(\bm q+\bm\omega_q^\perp)\bm{\cdot x}}\,j\,\right]
 \\ \label{S1}
\Phi_{II}&= T\rho_p\left[\cos\Theta_p e^{i(\bm p+\bm\gamma_p^\perp)\bm{\cdot x}}+\sin\Theta_p e^{i(-\bm p+\bm\omega_p^\perp)\bm{\cdot x}}\,j\,\right]
\end{align}
with $R$ and $T$ complex constants and $\bm k,\,\bm q,\,\bm p,\,\bm\gamma^\perp_a$ and $\bm\omega^\perp_a$ real vectors for $a=k,\,q,\,p$. 
We also adopt real constants for $\Theta_a$ and real functions for $\rho_a$. An arbitrary vector $\bm v$ is decomposed as
\begin{equation}\label{S2}
 \bm v=\bm v^\parallel +\bm v^\perp,
\end{equation}
where the component of $\bm v$ parallel to $\bm k$ is  $\bm v^\parallel$,  and the component of $\bm v$ normal to $\bm k$ is $\bm v^\perp$.
We expect that the three terms of the wave function (\ref{S1}) have identical properties, or describe particles of the same type. We choose the
time-dependent solution (\ref{A7}), where $E=0$. From (\ref{T7}), we obtain several energy relations
\begin{equation}\label{S3}
 \frac{2m\mathcal{E}}{\hbar^2}=\frac{|\bm \gamma_k|^2-|\bm\omega_k|^2}{2}=\frac{|\bm \gamma_q|^2-|\bm\omega_q|^2}{2},\qquad\qquad
\frac{2m(\mathcal{E}-V_0)}{\hbar^2}=\frac{|\bm \gamma_p|^2-|\bm\omega_p|^2}{2}
\end{equation}
and also that
\begin{equation}\label{S4}
 |\bm\alpha_a|^2=|\bm a|^2+\frac{|\bm \gamma_a|^2+|\bm\omega_a|^2}{2}\qquad\mbox{where}\qquad a=k,\,q,\,p.
\end{equation}
At the point of incidence $\bm x_0=(0,\,y_0,\,z_0)$, which we set to $\bm x_0=(0,\,0,\,0)$ without loss of
generality, we consider the continuity of the wave function, 
\begin{align}\label{S5}
\Phi_I (\bm x_0)=\Phi_{II}(\bm x_0)\qquad&\Rightarrow\qquad 
\left\{
\begin{array}{ll}
\rho_k(0)\cos\Theta_k^{(0)}+R\rho_q(0)\cos\Theta_q^{(0)}=&T\rho_p(0)\cos\Theta_p^{(0)}\\
\rho_k(0)\sin\Theta_k^{(0)}\,+R\rho_q(0)\sin\Theta_q^{(0)}=&T\rho_p(0)\sin\Theta_p^{(0)}
\end{array}
\right.
\\ \label{S8}
\bm\nabla\Phi_I^\parallel (\bm x_0)=\bm\nabla\Phi_{II}^\parallel(\bm x_0)\qquad &\Rightarrow\qquad 
\left\{
\begin{array}{ll}
&\;\;\;\bm k\rho_k(0)\cos\Theta_k^{(0)}-\bm q R\rho_q(0)\cos\Theta_q^{(0)}=\bm p T\rho_p(0)\cos\Theta_p^{(0)}\\
&-\,\bm k\rho_k(0)\sin\Theta_k^{(0)}+\bm q R\rho_q(0)\sin\Theta_q^{(0)}=-\bm p T\rho_p(0)\sin\Theta_p^{(0)}.
\end{array}
\right.,
\end{align}
and thus
\begin{equation}\label{S9}
 |T|^2=\frac{|\bm k+\bm q|^2}{|\bm p+\bm q|^2}\left(\frac{\rho_k(0)}{\rho_p(0)}\right)^2\qquad\mbox{and}\qquad 
|R|^2=\frac{|\bm k-\bm p|^2}{|\bm p+\bm q|^2}\left(\frac{\rho_k(0)}{\rho_q(0)}\right)^2.
\end{equation}
The normal directions place further boundary conditions
\begin{equation}\label{S10}
\bm\nabla\Phi_I^\perp (\bm 0)=\bm\nabla\Phi_{II}^\perp(\bm 0)\qquad \Rightarrow\qquad 
\left\{
\begin{array}{cc}
&\bm\nabla\rho_k(0)\,K_k+\bm\nabla\rho_q(0)\,R\,K_q=\bm\nabla\rho_p(0)\,T\,K_p\\
&\bm\gamma_k^\perp\,\rho_k(0)\,\cos\Theta_k^{(0)}+R\bm\gamma_q^\perp\,\rho_q(0)\,\cos\Theta_q^{(0)}=T\bm\gamma_p^\perp\,\rho_p(0)\,\cos\Theta_p^{(0)}\\
&\bm\omega_k^\perp\,\rho_k(0)\,\sin\Theta_k^{(0)}+R\bm\omega_q^\perp\,\rho_q(0)\,\sin\Theta_q^{(0)}=T\bm\omega_p^\perp\,\rho_p(0)\,\sin\Theta_p^{(0)},
\end{array}
\right.
\end{equation}
where
\begin{equation}\label{S11}
 K_a=\cos\Theta_a^{(0)}+\sin\Theta_a^{(0)}\,j,\qquad\bm\nabla\rho_a(0)=\bm\alpha_a(A_a-B_a)\qquad\mbox{and}\qquad a=k,\,q,\,p.
\end{equation}
We propose a solution using the constraint
\begin{equation}\label{S12}
\sin^2\Theta_k^{(0)}=\sin^2\Theta_q^{(0)}=\sin^2\Theta_p^{(0)}.
\end{equation}
If $|\bm k|=|\bm q|$, we benefit from the identity
\begin{equation}\label{S13}
|\bm k|-|\bm q|R\,\frac{\rho_q(0)}{\rho_k(0)}=|\bm p|T\,\frac{\rho_p(0)}{\rho_k(0)},
\end{equation}
which is valid for the scattering of quantum particles in CQM. Thus, (\ref{S10}), (\ref{S12}) and (\ref{S13}) give rise to
\begin{equation}\label{S16}
 \frac{|\bm\gamma_q|}{|\bm\gamma_k|}=\frac{|\bm\omega_q|}{|\bm\omega_k|}=\frac{|\bm\nabla\rho_q(0)|}{|\bm\nabla\rho_k(0)|}=1\qquad\qquad
\frac{|\bm\gamma_p|}{|\bm\gamma_k|}=\frac{|\bm\omega_p|}{|\bm\omega_k|}=\frac{|\bm\nabla\rho_p(0)|}{|\bm\nabla\rho_k(0)|}=\frac{|\bm p|}{|\bm k|}.
\end{equation}
From (\ref{S3}-\ref{S4}) and (\ref{S16}) we finally obtain
\begin{equation}\label{S17}
\frac{|\bm\alpha_p|^2}{|\bm\alpha_k|^2}=\frac{|\bm p|^2}{|\bm k|^2}\qquad\qquad \frac{|\bm p|^2}{|\bm k|^2}=1-\frac{V_0}{\mathcal{E}}.
\end{equation}
Consequently, every parameter of the reflected and transmitted particles may be written in terms of the incident particle parameters. The 
results are similar to the CQM case, but the transmission and reflection coefficients have multiplying factors that are characteristic 
of quaternionic particles. These factors are related to the oscillation along the $\bm\nabla\rho$ direction; as discussed in
section \ref{F}, the quaternionic particle propagates along $\bm\nabla\Gamma$ and $\bm\nabla\Omega$, whereas 
along $\bm\nabla\rho$ and $\bm\nabla\Theta$ it only oscillates.

\section{CONCLUSION\label{C}}
In this article, we have presented a general method for solving the QSE (\ref{A1}). The results are complementary 
to the solution of the QSE with a right multiplying complex unit $i$ \cite{Giardino:2017qqp}. We have proven that there are 
time-dependent solutions for Schr\"odinger equation
(\ref{A1}), which are fundamentally different from the time-function of CQM. We have also developed a general method for solving
the time-independent solution, and we have obtained free particle solutions without a CQM counterpart. These solutions present oscillations
that are normal to the propagation direction. This result is unknown in CQM, and now we need to find more physical examples where
this kind of situation is found. Another interesting new feature is the influence of quaternionic constants on the energies of solutions. 
This is another quaternionic feature unknown in CQM, and other physical systems where this kink of behavior is found are also of interest for
further research.

In summary, we expect that these two articles will foster new developments in QQM. First of all, because the presented solutions are simple, 
and do not assume the anti-hermitian constraint that has been supposed for the QQM. Secondly, we have a method for finding new solutions, and
then every result of CQM is at risk of being studied using the quaternionic formalism deployed here.
We expect that more physical solutions may be found in the future, which may inspire either mathematical or physics investigations,
including experimental ones.

\section*{ACKNOWLEDGEMENTS}
Sergio Giardino is grateful for the hospitality at the Institute of Science and Technology of Unifesp in S\~ao Jos\'e dos Campos.

%
%
%
%

\bibliographystyle{unsrt} 
\bibliography{bib_qfree2}

\end{document}